\documentclass[pra,twocolumn,showpacs,amsmath,amssymb,superscriptaddress,aps]{revtex4}

\usepackage{graphicx}% Include figure files
\usepackage{amsmath}
\usepackage{mathrsfs}
\usepackage{todonotes}
\usepackage{bm}% bold math

\begin{document}

\title{Information and backaction due to phase contrast imaging measurements of cold atomic gases: beyond Gaussian states}

\author{Ebubechukwu O. Ilo-Okeke} %\email[]{ }
\affiliation{National Institute of Informatics, 2-1-2 Hitosubashi, Chiyoda-ku, Tokyo 101-8430, Japan.}
\affiliation{Department of Physics, School of Science, Federal University of Technology,  P. M. B. 1526, Owerri, Imo State 460001, Nigeria.}

\author{Tim Byrnes} %\email[]{}
\affiliation{New York University Shanghai, 1555 Century Ave., Pundong, Shanghai 200122, China}
\affiliation{NYU-ECNU Institute of Physics at NYU Shanghai, East China Normal University, Shanghai 200062, China}
\affiliation{National Institute of Informatics, 2-1-2 Hitosubashi, Chiyoda-ku, Tokyo 101-8430, Japan.} 
\affiliation{Department of Physics, New York University, New York, NY 10003,USA}

\date{\today}% It is always \today, today,
\begin{abstract}
We further examine a theory of phase contrast imaging (PCI) of cold atomic gases, first introduced by us in Phys. Rev. Lett. {\bf 112}, 233602 (2014).  We model the PCI measurement by directly calculating the entangled state between the light and the atoms due to the ac Stark shift, which induces a conditional phase shift on the light depending upon the atomic state. By interfering the light that passes through the BEC with the original light, one can obtain information of the atomic state at a single shot level.  We derive an exact expression for a measurement operator that embodies the information  obtained from PCI, as well as the back-action on the atomic state. By the use of exact expressions for the measurement process, we go beyond the continuous variables approximation such that the non-Gaussian regime can be accessed for both the measured state and the post-measurement state. 
Features such as the photon probability density, signal, signal variance, Fisher information, error of the measurement, and the backaction are calculated by applying the measurement operator to an atomic two spin state system. For an atomic state that is initially in a spin coherent state, we obtain analytical expression for these quantities.  There is an optimal atom-light interaction time that scales inversely proportional to the number of atoms, which maximizes the information readout. 
\end{abstract}

\maketitle
%
%%%%%%%%%%%%%%%%%%%%%%%%%%%%%%%%%%%%%%%%%%%%%%%%%%%%%%%%%%%%%%%%%%%%%
%%%%%%%%%%%%%%%%%%%%%%%%%%%%%%%%%%%%%%%%%%%%%%%%%%%%%%%%%%%%%%%%%%%%%
\section{INTRODUCTION}
%%%%%%%%%%%%%%%%%%%%%%%%%%%%%%%%%%%%%%%%%%%%%%%%%%%%%%%%%%%%%%%%%%%%%
%%%%%%%%%%%%%%%%%%%%%%%%%%%%%%%%%%%%%%%%%%%%%%%%%%%%%%%%%%%%%%%%%%%%%
 
\begin{figure}[t]
\includegraphics[width=\columnwidth]{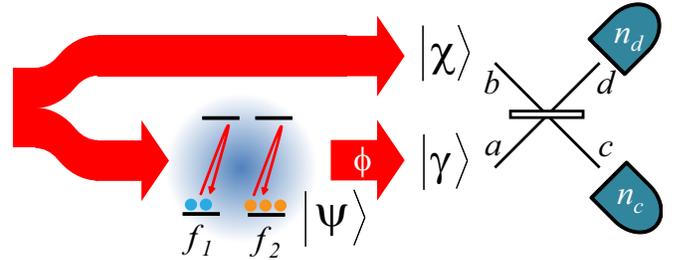}
\caption{Schematic configuration for the off-resonant light-matter interaction for phase contrast imaging considered in this paper. The incident light is detuned from atomic resonance realizing an ac Stark shift. The atoms in the state $ | \psi \rangle $ involving the ground state internal levels $ f_i $ with the incident light, and become entangled with the light. Every Fock state of the atoms produce a phase shift $ \hat{\phi} $.  Finally, the light which passes through the atoms $ | \gamma \rangle $ and the light which does not pass through $ | \chi \rangle $ are interfered and photons are counted to obtain a signal $ \cal S $.}
\label{fig:homodyne}
\end{figure}

Phase contrast imaging (PCI) \cite{zernike1942} is a powerful tool for performing non-destructive measurements, based on the observation of a phase shift of incident waves due to interaction with a probe object. It has been realized in a variety of configurations, involving light \cite{zernike1942}, X-rays \cite{bonse1965,momose1995} and electron beams \cite{spence1988} and found extensive use in biomedical imaging \cite{momose1996}, and structural imaging and studies of properties of nano devices and materials~\cite{smith1997}. In the context of measurement of atomic Bose-Einstein condensates (BEC), PCI \cite{andrews1996,bradley1997,higbie2005, kohnen2011,gajdacz2013} has been used to measure the properties of
ultracold atomic gases \cite{kohnen2011,gajdacz2013}, as well as small and dense atomic condensates \cite{andrews1996,bradley1997,higbie2005} \emph{in situ}.  The method has the advantage over other alternative methods such as absorptive \cite{anderson1995,andrews1997b} and fluorescence \cite{depue2000} imaging, as it can be applied repeatedly to the same
atomic sample without destroying it \cite{andrews1997,meppelink2010}. During the PCI measurement \cite{andrews1996,bradley1997,higbie2005}  a nonresonant detuned coherent light beam interacts via the ac Stark shift with the atoms causing the states of light to accumulate a phase shift as shown in Fig. \ref{fig:homodyne}. The phase shift of light is detected in a homodyne measurement from which the state of the atomic condensate is inferred. Such a measurement technique is a central component for readout in several applications proposing to use atomic BEC in quantum metrology \cite{shin2004,wang2005} and quantum information \cite{cory1997,byrnes2012,byrnes2015}.

%%%%%%%%%%%%%%%%%%%%%%%%%%%%%%%%%%%%%%%%%%%%%%%%%%%%%%%%%%%%%%%%%%%%%%%%%%%%%%%%%%%%%%
% FIGURE 1 INSERT BEGINS
%%%%%%%%%%%%%%%%%%%%%%%%%%%%%%%%%%%%%%%%%%%%%%%%%%%%%%%%%%%%%%%%%%%%%%%%%%%%%%%%%%%%%%

We previously developed a theory of PCI measurements at the single shot level in Ref. \cite{ilookeke2014}.  The basic idea of the method is to first explicitly calculate the entangled state between the atoms and the light induced by the ac Stark shift.  Projecting the photons on the interference basis, we were able to relate the difference of the photon counts to the average $ S^z $ spin of the BEC by a simple expression. The method both yields the amount of information about the quantum state that can be obtained, and the backaction on the state. In this paper, we give an extended treatment of the single shot PCI theory developed in Ref. \cite{ilookeke2014}.  We derive expression for the measurement operator that embodies both the information and  the backaction due to the measurement.  We show that the backaction  due to the measurement scrambles and shifts the relative phase of the atomic condensates, but is controllable by the degree of atom-light interaction. We demonstrate that the measurement has the best performance for the dimensionless atom-light interaction time being of the order $\sim 1/N$ where $N$ is the total number of atoms. Our findings complements earlier works that studied the limits on sensitivity \cite{lye2003,hope2004,hope2005} of nondestructive measurement, and the backaction \cite{leonhardt1999,dalvit2002,szigeti2009} due to PCI measurement. In particular, Refs. \cite{leonhardt1999,dalvit2002,szigeti2009} calculated the backaction of PCI measurement in a continuous measurement model by deriving an effective master equation, in which the degrees of freedom of probe is traced out.  Our work in contrast to these works are at the single-shot level, which directly shows how much information and backaction occurs with each PCI measurement. 

In addition to the above works, we note that Refs. \cite{sorensen1998,hald1999,hammerer2010} studied interactions between atomic ensemble and coherent light in a polarization measurement that is related to PCI. These works used the spectral density of the probe beam that has interacted with an atomic ensemble to estimate the degree of squeezing and entanglement resulting in the states of atomic ensemble. Our work is distinct from such works where the spin ensembles are treated in the continuous variables (CV) approximation. In this approach, the spins are considered to be polarized in a particular spin direction (say $S^x$), and the remaining total spin degrees of freedom ($S^y,S^z$) are used as the quadrature variables \cite{hammerer2010}.  In the approach presented in this paper, this approximation is not performed and thereby we are able to go beyond the CV regime.  For example, our constructed measurement operator can be applied to any state, not only spin coherent states in the vicinity of  polarized states.  Also it predicts the evolution of an initial spin coherent state into highly non-Gaussian states.  Such result obviously cannot be obtained in the CV approximations used in these works~\cite{sorensen1998,hald1999,hammerer2010}. Additionally, we aim to estimate the optimum scaling of atom-light interactions such that the measurement of the atomic condensates or ensemble is still minimally-destructive. We show that to calculate this one requires a beyond-CV analysis, as there is a fidelity loss that is not present in a CV approximation. These results are fundamental in nature and very important for applications using atomic condensates or ensembles and PCI in quantum information and metrology. 

The remainder of the paper is organized as follows. In Sec. \ref{sec:dipersiveimaging} we describe the model of PCI measurement that we use in this paper and Ref. \cite{ilookeke2014}, and derive an expression for the PCI measurement operator. We then analyze the information that is extracted from the PCI measurements using this operator in Sec. \ref{sec:information}.  This involves calculating the photon probability density (Sec. \ref{sec:probabilitydensity}), signal (Sec. \ref{sec:signal}), quality of the esimate (Sec. \ref{sec:optimisation}), and Fisher information (Sec. \ref{sec:fisher}).   We then examine the backaction due to the PCI measurement in Sec. \ref{sec:backaction}.  We give some experimental estimates of parameters in Sec. \ref{sec:exp}.  Finally, our conclusions are given in Sec. \ref{sec:summary}.

%%%%%%%%%%%%%%%%%%%%%%%%%%%%%%%%%%%%%%%%%%%%%%%%%%%%%%%%%%%%%%%%%%%%%
%%%%%%%%%%%%%%%%%%%%%%%%%%%%%%%%%%%%%%%%%%%%%%%%%%%%%%%%%%%%%%%%%%%%%
\section{PHASE CONTRAST IMAGING MODEL}
\label{sec:dipersiveimaging}
%%%%%%%%%%%%%%%%%%%%%%%%%%%%%%%%%%%%%%%%%%%%%%%%%%%%%%%%%%%%%%%%%%%%%
%%%%%%%%%%%%%%%%%%%%%%%%%%%%%%%%%%%%%%%%%%%%%%%%%%%%%%%%%%%%%%%%%%%%%
%
We consider an $ M $ component atomic Bose-Einstein condensate (BEC) confined in a trapping potential and is interacting with a light beam detuned from atomic resonant transition. The light beam couples the ground state to an excited state via an ac Stark shift as shown in Fig. \ref{fig:homodyne}. Assuming that the population of the atoms found in the excited state is negligibly small due to a large detuning, the excited state may be eliminated.  The effective interaction ac Stark shift Hamiltonian may be written (see the supplementary material of Ref. \cite{ilookeke2014})
\begin{equation}
\label{eq:bck01}
H = - \hbar \sum_{k=1}^M g_k \hat{n}_k  a^\dagger a,
\end{equation}
where $ \hat{n}_k = f_k^\dagger f_k $ is the number operator, $f_k$ are bosonic atomic annihilation operators that act on vacuum to destroy an atom on level $ k $ ($[ f_{k'}, f_k^\dagger] = \delta_{k k'} $), $g_k$ is the strength of atom-light interaction~\cite{ilookeke2014}, and $a$ is the light field operator that acts on the vacuum to destroy a photon ($[a,a^\dagger] = 1 $). Although we assume that the atoms are in a BEC, our theory equally applies to cold atom ensembles as long as the ac Stark shift acts symmetrically on all atoms, and the initial atomic state is symmetric under particle exchange.  The main difference formally is that the number operators for each level would be written $ \hat{n}_k = \sum_i | k,i \rangle \langle k,i | $ instead of bosonic operators, where $ |k,i \rangle $ denotes the $i$th atom in the ensemble in its $k$th atomic state.  As long as the atoms remain in a symmetric superposition state at all times, they are equivalent to the bosonic system.

We assume that the state of the BEC is 
\begin{equation}
| \psi \rangle  = \sum_{n_1 n_2\dots n_M} \psi_{n_1 n_2\dots n_M}   | n_1 n_2 \dots n_M \rangle,
\label{generalstate}
\end{equation} 
where $ | n_1 n_2 \dots n_M \rangle = | n_1 \rangle \otimes | n_2 \rangle  \otimes\dots \otimes| n_M \rangle $ and the Fock states of the BEC are 
\begin{align}
| n_k \rangle = \frac{1}{\sqrt{n_k!}} (f_k^\dagger)^{n_k} | 0 \rangle  .
\end{align}
The light field is initially in a coherent state which as suggested by Fig. \ref{fig:homodyne} is split into two components, light which passes through the BEC with amplitude $ \gamma $ and light which does not pass through the BEC of amplitude $ \chi $:
\begin{align}
\left|\gamma \right> & = e^{-\frac{|\gamma |^2}{2}} e^{\gamma a^\dagger } \left|0\right>, \nonumber \\
\left|\chi \right> & = e^{-\frac{|\chi |^2}{2}} e^{\chi b^\dagger }\left|0\right>,
\label{eq:bck03}
\end{align} 
where $ a $ ($ b $) is the photon annihilation operator for the light mode which does (not) pass through the BEC respectively.  
During the atom-light interaction, the Hamiltonian Eq. (\ref{eq:bck01}) entangles some of the light with the atomic condensate. The resulting state of the of the atom-light system is 
\begin{align}
& e^{-i H \tau/\hbar}  \left|\gamma\right> \left|\chi\right> | \psi \rangle \rightarrow \nonumber \\
&  \sum_{n_1 n_2\dots n_M} \psi_{n_1 n_2\dots n_M}  | \gamma e^{i\sum_k g_k n_k \tau}  \rangle |  \chi e^{i\phi_{\chi}}  \rangle  | n_1 n_2 \dots n_M \rangle ,
\label{eq:bck04}
\end{align}
In general, the light in the state $\left|\chi\right>$ will also pick up a phase $\phi_{\chi}$ after passing through the phase plate \cite{meppelink2010, higbie2005}, which is included in (\ref{eq:bck04}).  The atom-light interactions entangle the coherent states of atom and light, as the phase rotations of $ | \gamma e^{i\sum_k g_k n_k \tau}  \rangle $ will be different in general for each term in the expansion with $ | n_1 n_2 \dots n_M \rangle $.  The phase picked up by these photons contains some information about the state of the atoms. This information is accessed by interfering the light which passes through the BEC with the remaining light in a homodyne measurement as shown schematically in Fig. \ref{fig:homodyne}. Assuming a 50-50 beam splitter, the relationship to the new modes may be written 
\begin{align}
a^\dagger  & = \frac{c^\dagger + i d^\dagger}{\sqrt{2}} \nonumber, \\
b^\dagger  & = \frac{i c^\dagger + d^\dagger}{\sqrt{2}}. 
\end{align}
The state of atom-light system after light has passed through the beam splitter becomes 
\begin{align}
& e^{-\frac{|\gamma|^2 + |\chi|^2}{2}}  \sum_{n_1 n_2\dots n_M} \psi_{n_1 n_2\dots n_M} | n_1 n_2 \dots n_M \rangle
 \nonumber \\
& \times \exp[ \gamma e^{i\sum_k g_k n_k \tau}  (c^\dagger + i d^\dagger)/\sqrt{2} ]   \nonumber \\
& \times \exp[ \chi e^{i \phi_{\chi} } (ic^\dagger + d^\dagger)/\sqrt{2} ] \left|0\right>  .
\label{eq:bck05}
\end{align}
Finally, a photon number measurement is made in the number basis $n_c = c^\dagger c $ and $n_d = d^\dagger d$, which is achieved by projecting onto a particular $ n_c $ and $ n_d $ number state.  This gives
\begin{align}
& \frac{e^{-\frac{|\gamma|^2+ |\chi|^2}{2} }}{\sqrt{n_c!}\sqrt{n_d!}}
 \sum_{n_1 n_2\dots n_M} \psi_{n_1 n_2\dots n_M}  \nonumber \\
& \times \left[\frac{i}{\sqrt{2}}\left(-i\gamma e^{i \sum_k g_k n_k \tau }+ \chi e^{i\phi_{\chi}}\right)\right]^{n_c} \nonumber \\
& \times \left[\frac{1}{\sqrt{2}}\left(i\gamma e^{i \sum_k g_k n_k \tau } + \chi e^{i\phi_{\chi}}\right)\right]^{n_d} | n_1 n_2 \dots n_M \rangle . 
\label{finalstate}
\end{align}

We may now define a measurement operator $\hat{\mathcal{M}}_{n_c, n_d} $ which gives the total effect of the PCI measurement.  For a PCI measurement with photon counting outcome of $ n_c $ and $ n_d $, 
\begin{align}
\label{eq:bck06}
\hat{\mathcal{M}}_{n_c, n_d}  & \equiv \frac{e^{-\frac{|\gamma|^2+ |\chi|^2}{2} }}{\sqrt{n_c!}\sqrt{n_d!}}\left[\frac{i}{\sqrt{2}}\left(-i\gamma e^{i\hat{\phi} }+ \chi e^{i\phi_{\chi}}\right)\right]^{n_c} \nonumber \\
& \times \left[\frac{1}{\sqrt{2}}\left(i\gamma e^{i\hat{\phi}} + \chi e^{i\phi_{\chi}}\right)\right]^{n_d},
\end{align}
where 
\begin{align}
\hat{\phi}= \sum_{k=1}^M g_k \hat{n}_k \tau 
\end{align}
and we have reinstated $ n_k $ to an operator as it is yet to act on a state in (\ref{eq:bck06}).  
The operator $\hat{\mathcal{M}}_{n_c, n_d}$ describes the effect of measurement on the condensate induced by the PCI procedure. From the definition it is clear that the final state (\ref{finalstate}) is simply $ \hat{\mathcal{M}}_{n_c, n_d} | \psi \rangle $. In the following sections we will study various properties of this measurement operator.

\section{INFORMATION READOUT}
\label{sec:information}

In this section we examine the information that can be extracted from a PCI measurement.  We first obtain the 
probability distribution of the photon counts, which forms the foundation of the information that can 
be extracted. We apply the probability distribution to the analysis of two component spin-state to obtain an expression for the signal, equal to the difference between the photocounts on the two detectors. Also, we discuss the error and Fisher information of the PCI readout for a two component spin-state.

\subsection{Photon probability distribution}
\label{sec:probabilitydensity}

The information of quantum state of the BEC is inferred from the difference in photon counts $ n_c -n_d $.   To calculate this we require calculation of the joint probability $P(n_c,n_d)$ of counting $n_c$ and $n_d$ photons.  This can be calculated tracing over the projection of the state $\left|n_1 \dots n_M\right>$ on the atomic states
\begin{align}
P(n_c,n_d)  = \sum_{n_1 n_2\dots n_M}  | \langle  n_1 n_2 \dots n_M | \hat{\mathcal{M}}_{n_c,n_d}| \psi \rangle |^2 .
\label{eq:bck07}
\end{align}
Using (\ref{eq:bck06}), the expression for the probability $P(n_c,n_d)$ is written explicitly as
\begin{align}
P(n_c,n_d)  = \frac{e^{-{|\gamma|^2 - |\chi|^2} }}{{n_c!}{n_d!}} \sum_{n_1 n_2\dots n_M} |  \psi_{n_1 n_2\dots n_M} |^2
{\cal Y}_{n_1 n_2\dots n_M} 
\label{eq:bck08}
\end{align}
where 
\begin{align}
{\cal Y}_{n_1 n_2\dots n_M}  & = \frac{1}{2^{n_c + n_d}}| -i\gamma e^{i \sum_k g_k n_k \tau  }+ \chi e^{i\phi_{\chi}} |^{2 n_c} \nonumber \\
& |  i\gamma e^{i \sum_k g_k n_k \tau } + \chi e^{i\phi_{\chi}} |^{2 n_d} ,
\label{functiony}
\end{align}

The exact expression for the probability given in (\ref{eq:bck08}) does not lend itself to an easy interpretation. To be able to understand the behavior of the probability, let us analyze (\ref{functiony}) in terms of the trigonometric functions
\begin{align}
{\cal Y}_{n_1 n_2\dots n_M} &= \left[\frac{|\chi |^2 + |\gamma|^2}{2} - |\chi \gamma| \sin(\phi_{\chi} - \sum_k g_k n_k \tau) \right]^{n_c} \nonumber \\
&\times \left[\frac{|\chi |^2 + |\gamma|^2}{2} + |\chi \gamma| \sin(\phi_{\chi} -  \sum_k g_k n_k \tau) \right]^{n_d}  ,
\label{eq:bck09}
\end{align}
where we have absorbed the relative phase between $ \gamma $ and $ \chi $ into $ \phi_{\chi} $. Assuming that the total number of atoms in the BEC is $ N $
\begin{align}
\sum_{k=1}^M n_k = N , 
\end{align}
we may consider that the phase term $ \sum_k g_k n_k $ in (\ref{eq:bck09}) produces an average phase offset
\begin{align}
\phi_\gamma = \tau \sum_k g_k \frac{N}{M} . 
\end{align}
The relative phase around this average then depends upon the particular $ n_k $ configuration, and we may define
\begin{align}
\phi_r = \tau \sum_k g_k (n_k - \frac{N}{M}) .  
\end{align}
From (\ref{eq:bck09}) we may already observe that the typical timescales that the PCI regime will work with is 
\begin{align}
g_k \tau N \sim 1 .
\label{typicaltime}
\end{align}
The fact that this is the optimal timescale will be derived in more precisely in the following sections.

For  $n_c, n_d \gg 1$, and requiring the total number of photons ($n_d + n_c $) in the measurement be greater than the relative number of photons $(n_d - n_c)$ [i.e. $\tfrac{2|\chi||\gamma|}{|\chi|^2 + |\gamma|^2} (n_d + n_c) > |n_d - n_c|$], the function $ {\cal Y} $ can be approximated as 
\begin{align}
{\cal Y}_{n_1 n_2\dots n_M}   & = n_c^{n_c} n_d^{n_d} \left( \frac{|\gamma|^2 + |\chi|^2}{n_c + n_d}\right)^{n_c + n_d}  
e^{- \frac{(\phi_r - \bar{\phi})^2}{2 \sigma^2} },
\label{eq:bck10}
\end{align}
where 
\begin{align}
\bar{\phi} &=  \phi_{\chi} -\phi_\gamma  - \arcsin\left(\frac{(|\chi|^2 + |\gamma|^2)(n_d - n_c)}{2| \chi \gamma|(n_d + n_c)} \right) ,\\
\sigma^2 &=   \frac{4n_c n_d}{(n_d + n_c)} \left\{\left[\frac{2 (n_d +n_c)  |\chi \gamma|}{|\chi|^2 + |\gamma|^2}\right]^2 - (n_d - n_c)^2\right\}^{-1}.
\label{eq:bck11}
\end{align}

It is immediately evident that special interesting cases occur. For $|\gamma| \ll |\chi|$, only very small amount of the light beam pass through the atomic condensate. This limit corresponds to the current experimental realisation where the photon flux through the atomic BEC is small in comparison with the photon flux that are not scattered by the atomic condensate. Another interesting limit is $|\gamma| = |\chi|$, where the photon flux through the BEC is equal to the photon flux not scattered by the condensate. The scenario $|\gamma| = |\chi|$ can be realised experimentally by placing the condensate in one arm of Michelson interferometer.

We may now see how the photons become correlated with the quantum state by examining the peak of the distribution (\ref{eq:bck10}).  The maximum of the Gaussian distribution is located at
\begin{align}
n_c - n_d & = 2| \chi \gamma | \left(\frac{n_c + n_d}{|\chi|^2 + |\gamma|^2} \right) \sin (\phi_r - \phi_{\chi} + \phi_{\gamma} ) \nonumber \\
& \approx 2| \chi \gamma |  \sin (\phi_r - \phi_{\chi} + \phi_{\gamma} ) .
\label{generalncnd}
\end{align}
Where we have used the conservation of the number of photons such that the total number of photons that are detected is the same as that in the initial light. This relation suggests that up to the constant phase factors $ \phi_{\chi} $ and $  \phi_{\gamma} $, the photon count difference can be related to $ \phi_r $, the relative phase which depends upon the state distribution.

\subsection{Application to two component spin-state}
\label{sec:twocomponentatom}

Let us now specialize to the case where there are only two hyperfine ground states that the atoms occupy $ M = 2 $, and they are in spin coherent state
\begin{align}
| \psi \rangle & = \left.\left|\alpha_0,\beta_0\right>\right> \equiv \frac{1}{\sqrt{N!}}\left(\alpha_0 f^\dagger_1 + \beta_0 
f^\dagger_2\right)^N\left|0\right> \nonumber \\
& = \sum_{n_1=0}^N \sqrt{{ N \choose n_1}} \alpha_0^{n_1} \beta_0^{N-n_1} | n_1 N-n_1 \rangle . 
\label{eq:bck02}
\end{align} 
In this case the probability distribution is
\begin{align}
&  P(n_c,n_d)  = \frac{e^{-{|\gamma|^2 - |\chi|^2} }}{{n_c!}{n_d!}}  \nonumber \\
& \times \sum_{n_1=0}^N { N \choose n_1} |\alpha_0|^{2 n_1} |\beta_0|^{2N-2n_1} {\cal Y}_{n_1 N-n_1},  
\label{twocompprob}
\end{align}
where $ {\cal Y}_{n_1  N-n_1} $ is given in (\ref{eq:bck10}), and 
\begin{align}
\phi_\gamma & = g  \tau N,  \nonumber \\
\phi_r & = G \tau (2 n_1-N),  
\end{align}
with $ G = (g_1 - g_2)/2  $ and $ g = (g_1 + g_2)/2  $. From (\ref{typicaltime}) the typical interactions times that the PCI measurement will work in is 
\begin{align}
G \tau N \sim g \tau N \sim 1 . 
\end{align}

With the use of Stirling's approximation the binomial function in the expression for the probability is simplified as 
\begin{align}
\label{eq:bck12}
{ N \choose n_1}&  |\alpha_0|^{2 n_1} |\beta_0|^{2N-2n_1}  \approx \frac{1}{\sqrt{2\pi N |\alpha_0|^2 |\beta_0|^2}} \nonumber \\
&  \times \exp\left[ -\frac{N}{2 |\alpha_0|^2 |\beta_0|^2} \left(\frac{n_1}{N}  - \frac{|\alpha_0|^2 -|\beta_0|^2  +1}{2} \right)^2 \right] .
\end{align}
Replacing the sum over $n_1$ in (\ref{twocompprob}) by an integral and evaluating, we write the expression for the probability as 
\begin{equation}
\label{eq:bck13}
\begin{split}
 P(n_c,n_d)& = \frac{ n_c^{n_c} n_d^{n_d} \sigma  e^{-|\chi|^2-|\gamma|^2}}{n_c!n_d!}\left( \frac{|\gamma|^2 + |\chi|^2}{n_c + n_d}\right)^{n_c + n_d}  \\
& \times \frac{\exp \left[ -\frac{N^2}{2(G^2 \tau^2 N \sin^2 \theta_0  + \sigma^2)}
\left( \frac{\bar{\phi}}{2 G N \tau} -  \frac{\cos \theta_0}{2}\right)^2\right] }{\sqrt{G^2 \tau^2  N \sin^2 \theta_0   + \sigma^2}},
\end{split}
\end{equation}
where we have used a standard parametrization of the state on the Bloch sphere
\begin{equation}
\begin{split}
\alpha_0 & = e^{-i \varphi_0/2} \cos (\theta_0/2), \nonumber \\
\beta_0 & = e^{i \varphi_0/2} \sin (\theta_0/2) .
\end{split} 
\end{equation}

Finally using Stirling's approximation in (\ref{eq:bck13}) results in the following simplified expression for the probability density
\begin{align}
\label{eq:bck14}
& P(n_c, n_d) = \sigma \left(\frac{|\gamma|^2 + |\chi|^2}{n_c + n_d}\right)^{n_c + n_d} e^{(n_c + n_d -|\gamma|^2 - |\chi|^2)} \nonumber \\
& \times \frac{
\exp\left[ -\frac{N^2}{2(G^2 \tau^2 N\sin^2 \theta_0  + \sigma^2)}
\left( \frac{\bar{\phi}}{2 G N\tau} -  \frac{\cos \theta_0}{2}\right)^2\right] }{\sqrt{4\pi^2(G^2 \tau^2 N\sin^2 \theta_0  + \sigma^2)n_cn_d}}.
\end{align}

% %%%%%%%%%%%%%%%%%%%%%%%%%%%%%%%%%%%%%%%%%%%%%%%%%%%%%%%%%%%%%%%%%%%%%%%%%%%%%%%%%%%%%%
% % FIGURE PROBABILITY DENSITY INSERT BEGINS
% %%%%%%%%%%%%%%%%%%%%%%%%%%%%%%%%%%%%%%%%%%%%%%%%%%%%%%%%%%%%%%%%%%%%%%%%%%%%%%%%%%%%%%
\begin{figure}[t]
\includegraphics[width=0.8\columnwidth]{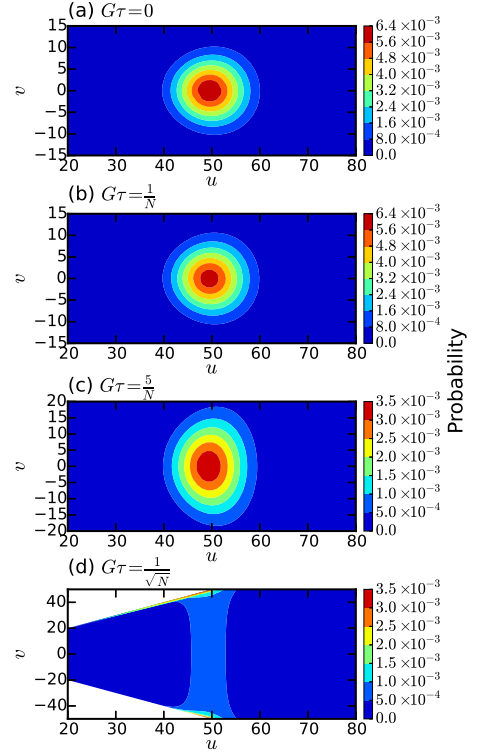}
\caption{(Color online). The PCI photon count probability density (\ref{eq:bck08}) in terms of relative and average photon numbers $u$, $v$. The parameters for the figures are $N = 1000$, $|\gamma|^2 = |\chi|^2 = 50$, $\theta_0  = \frac{\pi}{2} $, $\phi_{\gamma}  = \phi_{\chi} = 0$, and the atom-light interaction time $G\tau$ is as shown in each figure.   }
\label{fig:probability}
\end{figure}

The form of Eq.~(\ref{eq:bck14}) suggests that probability density be written in terms of relative photon number which by the way corresponds to what one expects in an experiment. Making a change of basis into the relative coordinates
\begin{align}
u & = \frac{n_c + n_d}{2}, \nonumber \\
v & = \frac{n_d - n_c}{2},
\end{align}
the probability becomes
\begin{align}
\label{eq:bck15}
&P(u, v)  = \sigma \left(\frac{|\gamma|^2 + |\chi|^2}{2u}\right)^{2u} e^{(2u -|\gamma|^2 - |\chi|^2)}\nonumber \\
& \times \frac{\exp\left[ -\frac{N^2}{2(G^2 \tau^2 N\sin^2 \theta_0  + \sigma^2)}
\left( \frac{\bar{\phi}}{2 G N\tau} -  \frac{\cos \theta_0}{2}\right)^2\right]  }{\sqrt{\pi^2(G^2 \tau^2 N\sin^2 \theta_0   + \sigma^2)(u^2 - v^2)}},
\end{align}
where 
\begin{equation}
\label{eq:bck16}
\begin{split}
\bar{\phi} &= \phi_{\chi} - \phi_{\gamma} -\arcsin\left(\frac{v(| \chi|^2 + |\gamma|^2)}{2 u| \chi \gamma|}\right) ,\\
\sigma^2 &=  \frac{(u^2 - v^2)(| \chi|^2 + |\gamma|^2)^2 }{2 u (4 u^2  | \chi \gamma|^2  - v^2 (| \chi|^2 + |\gamma|^2)^2)}.
\end{split}
\end{equation}
The most dominant contribution to the probability density comes from the points around the maximum of the function 
\begin{align}
P(u)  & = \left(\frac{|\gamma|^2 + |\chi|^2}{2u}\right)^{2u} e^{[2u -|\gamma|^2 - |\chi|^2]}\frac{1}{\sqrt{\pi u}} \nonumber \\
& \approx 
\sqrt{\frac{2}{\pi (|\gamma|^2 + |\chi|^2)}} e^{-\frac{(2u - |\gamma|^2 - |\chi|^2)^2}{2(|\gamma|^2 + |\chi|^2)}}.
\label{eq:bck17}
\end{align}

Expanding $P(u,v)$ around $ u= \frac{|\chi|^2 + |\gamma|^2}{2} $, we may write it as product of two functions that are properly normalized to unity
\begin{align}
P(u,v) \approx P(u) P(v),
\label{approxprob}
\end{align}
where
\begin{align}
& P(v) =   \sigma \sqrt{\frac{2(|\gamma|^2 + |\chi|^2)}{\pi((|\gamma|^2 + |\chi|^2)^2 - 4v^2)}} \nonumber \\
& \times \frac{\exp\left[ -\frac{N^2}{2( G^2 \tau^2 N\sin^2 \theta_0  + \sigma^2)}
\left( \frac{\bar{\phi}}{2 G N\tau} -  \frac{\cos \theta_0 }{2}\right)^2\right]  }{\sqrt{G^2 \tau^2  N\sin^2 \theta_0   + \sigma^2}}.
\label{pvfunc}
\end{align}

From (\ref{eq:bck17}) and (\ref{pvfunc}), the averages of the probability distribution are
\begin{align}
\langle u \rangle \approx \frac{|\gamma|^2 + |\chi|^2}{2},
\end{align}
and
\begin{align}
\langle v  \rangle \approx | \chi \gamma| \sin \Phi, 
\end{align}
while the variances of probability densities are approximately 
\begin{align}
(\Delta u)^2 \approx \frac{|\gamma|^2 + |\chi|^2}{2},
\end{align}
and 
\begin{align}
(\Delta v)^2 \approx &  \frac{G^2 \tau^2 N \sin^2 \theta_0  + \sigma^2 }{2(|\gamma|^2 + |\chi|^2)\sigma^2} \nonumber \\
& \times \left((|\gamma|^2 + |\chi|^2)^2- 4| \chi \gamma|^2 \sin^2 \Phi  \right),
\end{align}
respectively,  
where 
\begin{align}
\Phi = \phi_{\chi} - \phi_{\gamma}  - G\tau N \cos \theta_0  .  
\end{align}
The average and variance of the probability density with $u$ is independent of the atom-light interactions.  The probability density with $ v $ on the other hand depends on interaction time $G\tau$ as shown in Fig.~\ref{fig:probability}. For instance, at small values of $G\tau \ll 1/\sqrt{N}$ shown in Fig. \ref{fig:probability}(a)(b) the width of the relative probability density along $v$ is roughly $\sim \Delta u $.  Fig. \ref{fig:probability}(c) shows that increasing atom-light interaction time causes the width of the probability density along $v$ to grows linearly at a rate proportional to $G\tau$, 
\begin{align}
(\Delta v)^2 \approx \frac{(|\gamma|^2 + |\chi|^2)}{2\sigma^2}{(G^2 \tau^2 N\sin^2 \theta_0    + \sigma^2)}, 
\end{align}
while the width along $u$  remains the same. At large $G\tau \sim \tfrac{1}{\sqrt{N}}$, the width of the relative probability density along $v$ is of the order $(\Delta u)^2$ as can be seen in Fig. \ref{fig:probability}(d). We note that the anomalous features in Fig. \ref{fig:probability}(d) arise because $ v $ is only defined on the interval $ [-u,u] $.  For $v$ close to $\pm u$, the exponential term dependent on $v$ (i.e. $\bar{\phi})$ in (\ref{eq:bck15}) becomes small but finite. However, the amplitude approaches infinity such that the product remains finite and large in comparison to values of $v < |u|$. This explains the edge effect observed in Fig. \ref{fig:probability}(d) compared to Fig. \ref{fig:probability}(a)-(c).

% 
% %%%%%%%%%%%%%%%%%%%%%%%%%%%%%%%%%%%%%%%%%%%%%%%%%%%%%%%%%%%%%%%%%%%%%%%%%%%%%%%%%%%%%%
% % FIGURE SIGNAL AND VARIANCE INSERT BEGINS
% %%%%%%%%%%%%%%%%%%%%%%%%%%%%%%%%%%%%%%%%%%%%%%%%%%%%%%%%%%%%%%%%%%%%%%%%%%%%%%%%%%%%%%
\begin{figure}[t]
\includegraphics[width=\columnwidth]{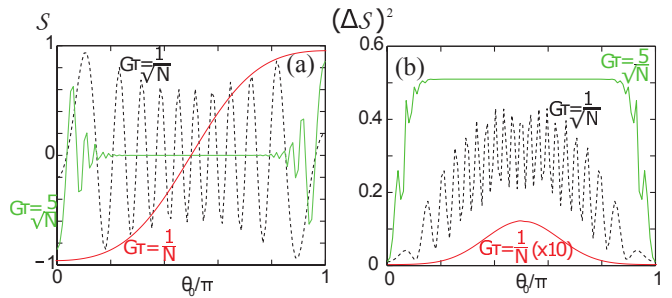}
\caption{The (a) signal and (b) variance of the PCI measurement of a spin coherent state with an initial state parameter $ \cos \theta_0 = | \alpha_0|^2 - | \beta_0|^2 $. The curve for $G\tau =\tfrac{1}{N}$ in (b) is magnified by a factor of 10 for visibility. Parameters used are $|\gamma|^2 = |\chi|^2 = 50$,  $N = 1000$, and $\phi_{\gamma} = \phi_{\chi} = 0$.  The dimensionless atom light interaction time $G\tau$ are as marked.\label{fig:mean} }
\end{figure}

%==========================================================
\subsection{Signal from photon counting}
\label{sec:signal}
%==========================================================

The signal $\mathcal{S}$ obtained from the measurement is calculated according to
\begin{align}
\mathcal{S} & \equiv \frac{\langle n_c \rangle - \langle n_d \rangle }{2|\chi \gamma|} 
= \frac{1}{2|\chi \gamma|}  \sum_{n_c, n_d} P(n_c,n_d) (n_c - n_d) ,
\label{signaldef}
\end{align}
where the normalization is taken for convenience such that the signal is a quantity of order unity, as suggested by (\ref{generalncnd}). 

In the general case (\ref{signaldef}) must be evaluated numerically. For specific states of the atomic system
it is possible to evaluate the expressions analytically.  For spin coherent states, we may use the approximate
probability distribution (\ref{pvfunc}) to evaluate
\begin{align}
\label{eq:bck22}
\mathcal{S} \approx & \frac{1}{2|\chi \gamma| } \int\limits_{-|\chi \gamma|}^{|\chi \gamma|} dv v P(v),\\
= & e^{-N G^2\tau^2 \sin^2 \theta_0 /2} \sin \Phi,
\label{signalresult}
\end{align}
where the equality in (\ref{signalresult}) is achieved by evaluating (\ref{eq:bck22}) in a complex plane using the steepest descent method.
We see that the signal has an oscillatory dependence to the relative population difference $\cos \theta_0 = |\alpha_0|^2 - |\beta_0|^2 $, showing that our theory captures the effect of the PCI measurement \cite{banaszek2001,sacchi2007,ilookeke2014}. We observe in addition that the signal $\mathcal{S}$ decays exponentially at large atom-light interaction times $G\tau \sim \tfrac{1}{\sqrt{N}}$. The signal decay arises because atom-light interaction causes each photon number state to evolve at different rate, and results in accumulation of a relative phase between different photon number states. Averaging over the many different photon number states, each number state evolving at different rate gives the exponentially decaying amplitude in (\ref{eq:bck22}). Similar effects have been observed in other systems involving $ S^z S^z $ interactions, where at equivalent times there is an ``oversqueezing'' effect and linear correlations are lost \cite{byrnes2013}. This suggests that to obtain the largest signal it is best to have times in the range $G\tau \sim \frac{1}{N}$.  This will be verified in the next section using different methods. 

Meanwhile, the variance of the measurement can be calculated similarly  
\begin{align}
 (\Delta \mathcal{S})^2 & \equiv   \frac{1}{4|\chi \gamma|^2}  \sum_{n_c, n_d} P(n_c,n_d) (n_c - n_d)^2  -  \mathcal{S}^2  
\label{variancedef} \\
& = \frac{| \chi |^2 + |\gamma|^2 }{4 | \chi \gamma|^2} + \frac 12 \left(1 - e^{-2 G^2\tau^2 N \sin^2 \theta_0 }\right) \nonumber \\
& - e^{-G^2\tau^2 N \sin^2 \theta_0 } (1 - e^{- G^2\tau^2 N \sin^2 \theta_0}) \sin^2 \Phi.
\label{eq:bck23}
\end{align}
The variance as written consists of contribution from two sources. The first term is the shot noise of the probe light, and remaining terms are due to the fluctuation in the atomic condensate.  At small values of the atom-light interaction time $ G\tau \sim 1/N$, the total variance increase and can be approximately be written as 
\begin{align}
\label{eq:bck24}
& (\Delta \mathcal{S})^2  \approx \frac{| \chi |^2 + |\gamma|^2 }{4 | \chi |^2  |\gamma|^2}  + \frac{\sin^2 \theta_0}{N} 
(G\tau N)^2 \cos^2\Phi .
\end{align}
By using a bright probe beam, one may reduce the shot noise fluctuations.  For sufficiently photon low shot noise, 
the fluctuations of the BEC can be observed. For a spin coherent state the variance  of $ S^z \equiv f^\dagger_1 f_1 - f^\dagger_2 f_2 $ is \cite{byrnes2012}
\begin{align}
\frac{(\Delta S^z)^2 }{N^2} = \frac{\sin^2 \theta_0}{N} ,
\end{align}
hence the PCI measurement can directly measure not only the average $  S^z $ spin but also the fluctuations.  
As the total variance $(\Delta \mathcal{S})^2$ oscillates as a function of the relative population difference in the atomic spin for a given $G\tau$, in such a variance estimate one must tune the phases such that the magnitude of the cosine is at a maximum, or equivalently the sine in (\ref{eq:bck22}) is at a minimum.  Thus the maximum variance measurement point is when the signal is at the minimum.  At longer atom-light interaction times $G\tau \sim  \frac{1}{\sqrt{N}}$, from (\ref{eq:bck23}) we see that the correlations
to the atomic state diminish in a similar way to the signal.  The PCI measurement degrades in this regime, hence for the variance interaction times $G\tau \sim  \frac{1}{N}$ is optimal.  

In Fig. \ref{fig:mean} we plot the signal and variance of the signal for typical experimental parameters, using the expression (\ref{signaldef}) and (\ref{variancedef}).  For times $ G \tau = 1/ N $ we see the expected behavior, where the signal oscillates with respect to angle $ \theta_0 $ to the $ S^z $-axis of the Bloch sphere. The variance also shows the expected behavior, where the maximum variance is seen when the magnitude of the signal is smallest.  Disregarding the shot noise which is small for the parameters chosen, (\ref{eq:bck24}) agrees with the form of the variance as plotted in Fig. \ref{fig:mean}(b).  Up until times $G\tau = \frac{1}{\sqrt{N}}$, we see that initially the period of the oscillations in the signal start to increase, as expected from (\ref{signalresult}). The variance has the general form of the $ \sin^2 \theta_0 $ envelope in (\ref{eq:bck24}) with oscillations due to the $ \cos^2 \Phi $.  However, for longer times $G\tau = \frac{5}{\sqrt{N}} $ the signal starts to deteriorate, due to the exponential dampening factor (\ref{signalresult}).  For these long times $ G \gtrsim 1/\sqrt{N} $, the variance starts to approach a constant value
\begin{align}
(\Delta \mathcal{S})^2 \rightarrow  \frac{| \chi |^2 + |\gamma|^2 }{4 | \chi |^2  |\gamma|^2}  + \frac{1}{2} ,
\label{largevariance}
\end{align}
which can be obtained by setting all the exponential factors in (\ref{eq:bck23}) to zero.  

We remark that in the limit $|\gamma| \ll |\chi|$, the observed signal is $\mathcal{S}_{|\gamma| \ll |\chi|} = |\gamma|\mathcal{S}$, and the variance is $(\Delta \mathcal{S}_{|\gamma| \ll |\chi|})^2 = |\gamma|^2(\Delta\mathcal{S})^2$. It is easily seen that the amplitude of the signal $\mathcal{S}_{|\gamma| \ll |\chi|}$ is as large as $|\gamma|$. On the other hand, the error of the PCI  measurement is dominated by contributions from the light that did not pass through the atoms $|\chi|$ (local oscillator) that is overlaid by the fluctuations in the atomic BEC that is of the order $|\gamma|^2$. The contributions due to the weak field $|\gamma|$ is vanishingly small $\sim (|\gamma|/|\chi|)^2$, and thus negligible.

% 
% %%%%%%%%%%%%%%%%%%%%%%%%%%%%%%%%%%%%%%%%%%%%%%%%%%%%%%%%%%%%%%%%%%%%%%%%%%%%%%%%%%%%%%
% % FIGURE SENSITIVITY INSERT BEGINS
% %%%%%%%%%%%%%%%%%%%%%%%%%%%%%%%%%%%%%%%%%%%%%%%%%%%%%%%%%%%%%%%%%%%%%%%%%%%%%%%%%%%%%%
\begin{figure}[t]
\begin{center}
\includegraphics[width=\columnwidth]{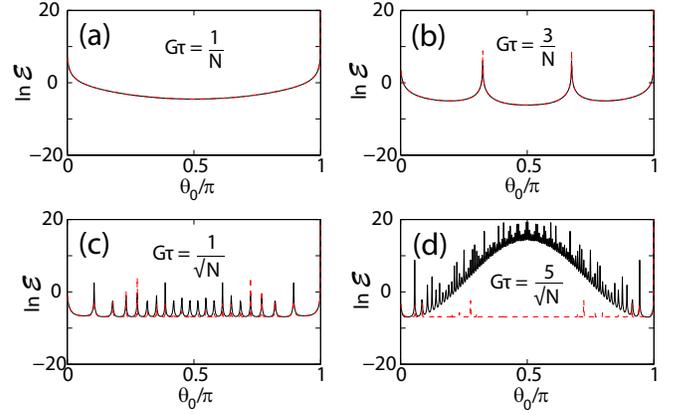}
\caption{(Color online) The error of the PCI measurement for an initial spin coherent state with Bloch sphere angle $\theta_0$. The interaction time is as shown in each figure. For (a)(b)(c) and (d) the solid line is (\ref{eq:bck26}) while the dashed line is (\ref{eq:bck27}). In all figures $\phi_{\chi} = \phi_{\gamma} = 0$.}\label{fig:sensitivity}
\end{center}
\end{figure}

%==========================================================
% Signal Sensitivity
%==========================================================
\subsection{Quality of the PCI measurement}
\label{sec:optimisation}
%==========================================================

The quality of PCI measurement in estimating the atomic spin, is quantified by the error propagation formula \cite{boixo2008,tikhonenkov2010}
\begin{equation}
\label{eq:bck25}
{\cal E} (\theta_0)  \equiv \frac{(\Delta \mathcal{S})^2 }{\left(\frac{\partial \mathcal{S}}{\partial\theta_0}\right)^{2}} .
\end{equation}
A good PCI measurement according to this measure has a small value of $ {\cal E} $, where the variance of the measurement is small and there is a strong correlation between the signal and the initial state. For zero interaction between the light and the atoms $ G \tau = 0 $,  the signal (\ref{signalresult}) has no dependence on $ \theta_0 $, thus $ {\cal E} $ is large and positive.

Using (\ref{signalresult}), $ {\cal E} $ may be written as 
\begin{equation}
\label{eq:bck26}
{\cal E} (\theta_0) =  \frac{(\Delta {\cal S})^2 e^{4N |\alpha_0 \beta_0 |^2 G^2\tau^2}}{\left[2GN \tau \left(2\cos\Phi\sin\theta_0 - G\tau\sin\Phi\sin 2\theta_0 \right)\right]^2}. 
\end{equation}
where $ (\Delta {\cal S})^2 $ is as given in (\ref{eq:bck23}).  Eq. (\ref{eq:bck26}) is plotted in Fig.~\ref{fig:sensitivity} as a function of $\theta_0$ for various interaction times $G\tau$. With increasing $G\tau$ in the regime $ 0< G\tau \lesssim 1/N $, the quality improves with the error $ {\cal E} $ generally reducing. At times $G\tau = 3/N$ some cusps develop due to the faster oscillation of the signal with $ \theta_0 $, which may be seen from the $ \Phi $ dependence in the denominator.  However, overall the error $ {\cal E} $ remains at a low level.  But, for $G\tau$ of the order of $1/\sqrt{N}$, more and more cusps develop degrading the quality of the PCI measurement. For $G\tau = 5/\sqrt{N} $ we observe a further degradation of the PCI quality with and overall increase of the error $ {\cal E} $.  As may be seen from (\ref{eq:bck26}), the error exponentially degrades, as the variance $ (\Delta {\cal S})^2 $ approaches a constant as seen from (\ref{largevariance}).

To understand this behavior, we may approximate (\ref{eq:bck26}) by expanding the exponential factors for small values of $G\tau \ll 1/\sqrt{N}$ to order $G^2\tau^2 $, giving
\begin{equation}
\label{eq:bck27}
{\cal E} (\theta_0)  \approx \frac{1}{N} + \frac{|\chi|^2+|\gamma|^2}{16 |\chi \gamma|^2  N^2G^2\tau^2\sin^2\theta_0\cos^2\Phi}.
\end{equation}
It is easily seen that the error (\ref{eq:bck27}) diverges as one approaches  $\theta_0 = 0,\pi$ which may be expected as these are the poles on the Bloch sphere, where with respect to the $ S^z $-axis, there is no variation with $ \theta_0 $. 
The lowest achievable error $ \cal E $ can be estimated by minimizing (\ref{eq:bck27}) with respect to $G\tau$ at a fixed $\alpha_0,\beta_0$, giving a criterion
\begin{equation}
\label{eq:bck28}
\tan [ G\tau N \cos \theta_0 -\phi_{\chi} + \phi_{\gamma} ] = \frac{1}{G\tau N\cos \theta_0 } . 
\end{equation}
For $ \phi_{\chi} =\phi_{\gamma} = 0 $ this has a solution $ G\tau N \cos \theta_0 \approx 0.86 $, which shows that to a reasonable estimate taking $ G\tau N \sim 1 $ will give close to optimum results. The scaling of the optimum coupling time has been previously obtained using a different means~\cite{ilookeke2014}.

% 
% %%%%%%%%%%%%%%%%%%%%%%%%%%%%%%%%%%%%%%%%%%%%%%%%%%%%%%%%%%%%%%%%%%%%%%%%%%%%%%%%%%%%%%
% % FIGURE Fisher INSERT BEGINS
% %%%%%%%%%%%%%%%%%%%%%%%%%%%%%%%%%%%%%%%%%%%%%%%%%%%%%%%%%%%%%%%%%%%%%%%%%%%%%%%%%%%%%%
\begin{figure}[t]
\begin{center}
\includegraphics[width=\columnwidth]{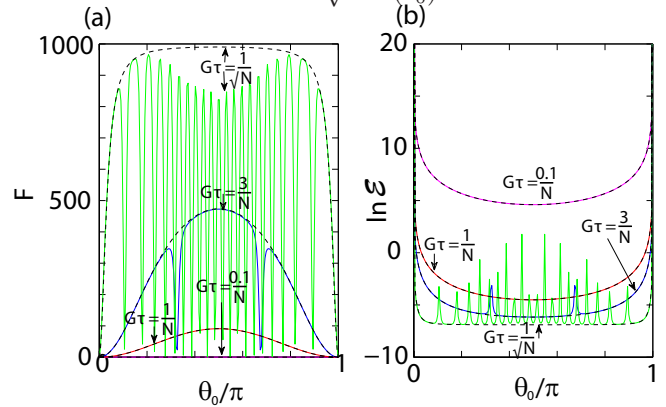}
\caption{(Color online). (a) Fisher information for a PCI measurement of a spin coherent state with Bloch sphere parameter $\theta_0$. (b) Error estimated from Fisher information as defined in (\ref{eq:bck31}). The parameters for the figures are $|\gamma|^2 = |\chi|^2 = 50$, $N = 1000$, and $ \phi_{\chi} = 0$. The atom-light interaction time $G\tau$ is as shown in the figure. For all curves, the solid line is the numerical curve and the dashed line is the theoretical curve.}\label{fig:fisher}
\end{center}
\end{figure}

\subsection{Fisher information}
\label{sec:fisher}

Another measure of the quality of the PCI measurement is the Fisher information defined as
\begin{equation}
\label{eq:bck30}
F(\theta_0) \equiv \sum_{n_c,n_d} {P(n_c,n_d)} \left(\frac{\partial \ln P(n_c,n_d)}{\partial\theta_0}\right)^2.
\end{equation}
This can be related to the error of PCI measurement through the Cram{\'e}r-Rao lower bound
\begin{equation}
\label{eq:bck31}
{\cal E} (\theta_0)  \geq \frac{1}{\sqrt{m F(\theta_0)}}.
\end{equation}
where $m$ is the number of independent repetitions of the experiment. For a given $m$, the Cram{\'e}r-Rao bound is minimum if the Fisher information $F(\theta_0)$ is large. Using the probability distribution (\ref{twocompprob}), we obtain an expression
\begin{equation}
\label{eq:bck32}
F(\theta_0) = \frac{ G^2 \tau^2 N^2 \sin^2\theta_0}{ G^2 \tau^2 N\sin^2\theta_0 + \sigma^2 },
\end{equation}
where $\sigma $ is 
\begin{equation}
\label{eq:add01}
\sigma^2 = \left\{4 |\chi\gamma|^2(|\chi|^2 + |\gamma|^2) \frac{1 - \sin^2\Phi}{(|\chi|^2 + |\gamma|^2)^2 - 4 |\chi\gamma|^2\sin^2\Phi} \right\}^{-1}.
\end{equation}
The Fisher information may also be written as 
\begin{widetext}
\begin{displaymath}
F(\theta_0) = \frac{N^2}{\tan^2(\theta_0/2)}\left\{\dfrac{\left(\sigma^2 + G^2 \tau^2 N \sin^2\theta_0\right)
\exp\left[\dfrac{4 G^2 \tau^2 \sin^4 (\theta_0/2)}{\sigma^2 + (N+1) G^2 \tau^2  \sin^2\theta_0}\right]}{\sqrt{\left(\sigma^2 + (N-1) G^2 \tau^2 \sin^2\theta_0\right) \left(\sigma^2 + (N+1) G^2 \tau^2 \sin^2\theta_0\right)}} -1 \right\} .
\end{displaymath}
\end{widetext}

We see that as $G\tau$ tends to zero, the Fisher information is zero meaning that no information can be inferred from the measurement. As such the sensitivity decreases (i.e. ${\cal E} (\theta_0)$ is infinite). However, as $G\tau$ tend to infinity, the Fisher information reaches a finite value of $N$, and one attains the best sensitivity which for $m = 1$ scales as $1/\sqrt{N}$. A comparison of the approximate result (\ref{eq:bck32}) with the numerical computation of (\ref{eq:bck30}) shows that with $G\tau$ the Fisher information does not increase indefinitely, and there is an optimum value $G\tau  \sim 1/N$ beyond which no significant information is gained by increasing the interaction time $G\tau$. This is because for $G\tau  > 1/N $ the Fisher information drops rapidly to zero for certain values of $\theta_0$, such that the singular points that are averaged out in the approximate expression starts to make a contribution. Thus we again see that the best PCI measurements are obtained for $ G \tau \sim 1/N $.

% 
% %%%%%%%%%%%%%%%%%%%%%%%%%%%%%%%%%%%%%%%%%%%%%%%%%%%%%%%%%%%%%%%%%%%%%%%%%%%%%%%%%%%%%%
% % FIGURE Q-FUNCTION INSERT BEGINS
% %%%%%%%%%%%%%%%%%%%%%%%%%%%%%%%%%%%%%%%%%%%%%%%%%%%%%%%%%%%%%%%%%%%%%%%%%%%%%%%%%%%%%%
\begin{figure}[h!]
\begin{center}
\includegraphics[width=\columnwidth]{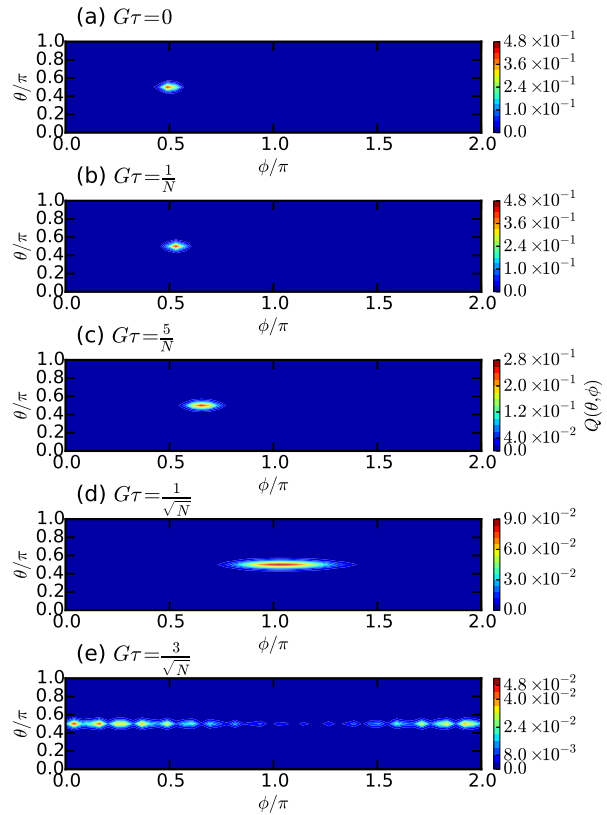}
\caption{(Color online). The \emph{Q}-function of the atomic BEC state after PCI imaging for various atom-light interaction times $G\tau$, as shown in the figures. The initial condition of the state of the atomic BEC is $\theta_0 = \pi/2$, $\varphi_0 = \pi/2$, and the parameters for the figures are $|\gamma|^2 = |\chi|^2 = 15$, $N = 300$, and $\phi_{\gamma} = \phi_{\chi} = 0$.}\label{fig:qfunction}
\end{center}
\end{figure}

%==========================================================
%Back-Action of Measurement
%==========================================================
\section{MEASUREMENT-INDUCED BACK-ACTION}
\label{sec:backaction}
%==========================================================

As we have seen in Sec. \ref{sec:dipersiveimaging}, the coupling the atomic state to light results in the photons to  accumulating a relative phase rotation. The detection of the phase shifts of the photons in a homodyne measurement causes the initial atomic state $ | \psi \rangle $ to make a transition to the state 
\begin{equation}
\label{eq:bck50}
\left|\psi_m\right> = \frac{\hat{\mathcal{M}}_{n_c,n_d}| \psi \rangle }{\sqrt{P(n_c,n_d)}}.
\end{equation}
with probability $P(n_c,n_d)$. This post-PCI measurement state has varying degrees of backaction depending upon the parameters chosen.  In this section we study the effect of the PCI measurement on the atomic quantum state at the single shot level.

In the ideal case, the PCI measurement produces a negligible backaction and the measurement is able to read out the state to 
a high fidelity.  Naturally, the laws of quantum mechanics imply that doing both perfectly is impossible.  However, as $ N $ becomes larger this scenario is asymptotically approached.  The information-disturbance tradeoff was calculated in Ref.  \cite{ilookeke2014} and was found to have a universal behavior.  To determine the effect of the backaction on the initial state $ | \psi \rangle $ it is instructive to plot the $Q$-distribution, which in our case is
\begin{align}
Q(\theta, \varphi) = & \frac{N + 1}{4\pi} \sum\limits_{n_c, n_d}P(n_c,n_d)\nonumber \\
& \times \left| \left<\left<\cos \tfrac{\theta}{2} e^{-i\varphi/2} ,\sin \tfrac{\theta}{2} e^{i\varphi/2} |\psi_m\right>\right.\right|^2.
\label{eq:bck51}
\end{align}
This is more representative than alternative measures of the backaction such as the fidelity (i.e. $ | \langle \psi | \psi_m \rangle |^2 $), as for many-particle states such as those that we deal here, there is an exponential suppression of the fidelity with the particle number.  For example, for two spin coherent states that deviate by an angle $ \delta \theta $ \cite{byrnes2012}, 
\begin{align}
\langle \langle \cos \frac{\theta}{2}, \sin  \frac{\theta}{2} | \cos \frac{\theta + \delta \theta }{2}, \sin  \frac{\theta+ \delta \theta }{2}  \rangle \rangle \approx e^{-N (\delta \theta)^2/8},
\end{align}
thus a fidelity becomes exponentially sensitive to small angular deviations on the Bloch sphere.  

For an initial spin coherent state $ |\psi \rangle = | \alpha_0, \beta_0 \rangle \rangle $, we may use similar approximations to that discussed in the previous section.  The \emph{Q}-function may be estimated analytically, giving 
\begin{align}
Q(\theta,\varphi) & = \frac{(N+1) \cos^{2N+1}(\frac{\theta}{2} - \frac{\theta_0}{2}) }{2^{3/2}\pi \sigma_Q \sqrt{N  \sin \theta \sin \theta_0} } e^{ -  \frac{(\varphi - \varphi_0 - (|\chi|^2 + |\gamma|^2) G\tau)^2}{2 \sigma_Q^2} },
\label{eq:bck52}
\end{align}
where 
\begin{align}
\sigma_Q^2 = \frac{2\cos^2  (\tfrac{\theta}{2} - \tfrac{\theta_0}{2})}{N \sin \theta \sin \theta_0} + \frac{2 G^2 \tau^2}{\sigma^2},
\end{align}
and $\sigma^2 $ is as given in (\ref{eq:add01}).
The analytical expression shows that the measurements affects only the phase $\varphi$ of the atomic condensates.  
In Fig. \ref{fig:qfunction} we directly calculate the $Q$ function numerically using (\ref{eq:bck51}), which shows the same general behavior, a broadening in the $ \varphi $ direction.  From the analytical form, we see the measurement causes a drift in the phase of the atomic condensate, $\varphi - \varphi_0 = ( |\chi|^2+ |\gamma|^2) G\tau$. This can be understood to be an average phase drift given by the light on the BEC. The other effect is the increase in the width of the relative phase to width $ \sigma_Q$.  Both effects are proportional to the atom-light interaction time, $G\tau$. Weak measurements with the interaction times $G\tau \ll 1/\sqrt{N}$ as described above,  tend to preserve the coherent properties of the state $\left|\psi_m\right>$ of atomic condensate, namely $\varphi \approx \varphi_0$ and $\Delta\varphi \sim 1/\sqrt{N}$. This is true provided that effect of photon statistics on the measurement is small $G\tau ( |\chi|^2 + | \gamma|^2 )\ll 1 $, as  seen in Figs. \ref{fig:qfunction}(a)(b). However, as the atom-light interaction time increases the relative phase of state $\left|\psi_m\right>$  is shifted by an amount proportional to the atom-light interaction time $G\tau$. The growth of $ \sigma_Q $ is due to a scrambling of the relative phase that causes loss of coherence as shown in Fig \ref{fig:qfunction}(c)(d)(e). This is a similar effect to phase diffusion due to atom-atom interaction, which causes each atomic  number state to evolve at different rate around the Bloch sphere,  thereby scrambling the phase \cite{lewenstein1996,javanainen1997}.

The fidelity of the states $\left|\psi\right>$ (\ref{eq:bck02}) and $\left|\psi_m\right>$ (\ref{eq:bck50})(not keeping the measurement record) is $F(\left|\psi\right>,\left|\psi_m\right>)  = \sum_{n_c,n_d}P(n_c,n_d)\left|\left<\psi|\psi_m\right>\right|^2 = F(\theta_0)$ where it easily shown that
\begin{equation}
\label{eq:add02}
F(\theta_0) = \frac{\sqrt{2}}{\sigma_F\sqrt{N}\sin\theta_0} e^{-\frac 12\frac{(|\chi|^2 + |\gamma|^2 )^2 G^2\tau^2}{\sigma_F^2} },
\end{equation}
where 
\begin{equation}
\label{eq:add03}
\sigma_F^2 = \frac{2}{N\sin^2\theta_0} + \frac{2G^2\tau^2}{\sigma^2},
\end{equation}
and $\sigma$ is as defined in (\ref{eq:add01}). Comparing (\ref{eq:add02}) and (\ref{eq:bck52}), it is evident that setting $\theta =\theta_0$ and $\phi = \phi_0$ in (\ref{eq:bck52}) and ignoring the normalisation constant gives the fidelity $F(\theta_0)$. For $G\tau = 0$, the fidelity is unity for all initial value as expected. However, the fidelity decreases with increasing values of $G\tau$  and is vanishingly small for all values of $\theta_0$ except for $\theta_0 = 0, \pi$. To understand this, we turn to the \emph{Q}-function where it is easily understood that for the atomic coherent state, the states $\theta_0 = 0, \pi$ is a well defined number state with zero phase and are not affected by the fluctuation in phase. In fact, whereas the \emph{Q}-function shows that the changes in the states of the atoms are due to distortion in the phase of the atomic state, the fidelity for a given initial amplitude $\theta_0$ measures the resemblance between the initial state and the final state. From the outset, it easily understood that the resemblance of the final state to the initial state is remarkably similar and the fidelity is close to unity if the phase is relatively unscrambled as discussed in the \emph{Q}-function section above. Hence, the fidelity indirectly measures the distortions in the phase of a given amplitude of atomic states, and the distortion proves to be roughly same amount for every initial amplitude $\theta_0$.

\section{Experimental parameter estimates} 
\label{sec:exp}

Lastly, we provide estimates for parameters appearing in our model. The interaction frequency  $G_j$ is $ G_j = \frac{\left|\left<g_j|d_j\cdot\varepsilon^*|e_j\right>\right|^2}{\hbar^2\Delta_j}\left<E^2(\mathbf{r})\right>$ ~\cite{ilookeke2014}, where the average field $\left<E^2(\mathbf{r})\right>$ in terms of the average intensity $I$ is $\left<E^2(\mathbf{r})\right> = \tfrac{2I}{\epsilon_0 c}$. For definiteness, we consider the $\mathrm{D}_1$ line transition of $^{87}\mathrm{Rb}$ atoms where only the $m_F= \pm1$ ($F=1$) states are occupied. For PCI imaging of $^{87}\mathrm{Rb}$ atomic BEC using $\sigma^+$ polarised laser light of average intensity $300\mu\mathrm{W/cm^2}$~\cite{higbie2005} and detuning of $212$~MHz~\cite{higbie2005}, $G= G_1 - G_{-1} =20 \times 10^3\,\mathrm{s}^{-1}$.  To estimate the interaction time $\tau$, we calculate the time taken by light to traverse the atomic cloud. In Ref~\cite{higbie2005}, PCI laser light was applied to BEC along the axis of tightest confinement. The radius of the cloud along this axis is $R_y = (\tfrac{2\mu}{m\omega_y^2})^{1/2}$, giving $\tau = 2R_y n_\mathrm{p}/c$, where $n_\mathrm{p}$ is the refractive index at the peak density of the atoms, $\mu$ is the chemical potential of the BEC, $\omega_y$ is the trap frequency along the axis of tightest confinement, and $m$ is the mass  of the atom. Thus the coupling strength $G\tau$ using parameters of Ref.~\cite{higbie2005} is $G\tau = 2 \times 10^{-7}$. Compared with $1/N = 2.5 \times 10^{-7}$ affirms that the experiment of Ref.~\cite{higbie2005} was performed in the  minimally-destructive regime.

To enter the non-Gaussian regime, we require coupling strength in the region of $G\tau \sim 1/\sqrt{N} $.  This thus requires a further increase of the dimensionless coupling $ G \tau $ by a factor $ \sim \sqrt{N} $. For small ensembles this may readily be achieved by increasing the laser intensity. Alternatively, cavities may be used to enhance the coupling. Enhancements of $ G $ by factors over $ 10^3 $ are readily obtainable~\cite{colombe2007,brennecke2007}, and hence the non-Gaussian regime should be reachable using current experimental technology.

%===============================================================
 \section{SUMMARY AND CONCLUSIONS \label{sec:summary}}
%===============================================================
In summary, we have presented a theory of single-shot phase contrast imaging of atomic Bose-Einstein condensates, extending upon the initial work presented in Ref. \cite{ilookeke2014}. We derived a measurement operator that fully describes the information obtained from the measurement, as well as the backaction due to the measurement.  Using the measurement operator, we calculated the probability density and its characteristic features such as the mean, standard deviation, estimation error and Fisher information. For the measurement to be described as non-destructive, we found that there is an optimum atom-light interaction time, which scales inversely with the population of atoms in the condensates, $G\tau \sim 1/N$. Beyond this atom-light interaction time $G\tau > 1/N$, the signal starts to deteriorate until no significant information can be inferred from the measurement. We showed using the \emph{Q}-function that the state of the atomic condensates suffers significant back-action due to the measurement for times $G\tau \gg 1/N$, but is fairly minimal for $G\tau \sim 1/N$. In particular, the back-action shifts and scrambles  the relative phase of atomic BEC states by an amount proportional to the atom-light interaction time, assuming that the photon statistics plays limited role $N \gg |\chi|^2 + |\gamma|^2 $ (i.e. the total number of atoms $N$ in the atomic condensate is greater than the combined average photon number $u_0$ used in the measurement). 

In this work we did not take into account the backaction resulting from dephasing due to residual absorption as done in Refs.~\cite{dalvit2002,leonhardt1999,szigeti2009}. Although light used in the measurement is far-detuned from atomic resonance transition, a small number of atoms are excited. These atoms decay by spontaneous emission, and do not return to the atomic condensate. Because the spontaneous decay process is uncontrolled and random, it leads to the heating of the atomic samples. Recently, it was found that narrow linewidth lasers are highly effective at suppressing the ac Stark shift scattering rate due to non-Markovian effects \cite{lone2015}, hence we expect that this can be reduced to a very low level in practice. We have also assumed that the particle number in the BEC is a constant $ N $, which may appear to be a strong assumption, as typically shot to shot the particle number in an experiment will vary.  However we point out that we consider a single-shot scenario where the number of atoms in the trap, including both the condensed and thermal fractions, is to a good approximation fixed to $ N $. As long as the light is applied to the atoms in a symmetric way, our theory applies to both BECs and ensembles, hence to first order we expect that both the condensed and uncondensed parts contribute in the same way. A full calculation taking these effects into account is left as future work.

\begin{acknowledgments}
T. B. thanks Vladan Vuletic for comments.  This work is supported by NTT Basic Research Laboratories, the Shanghai Research Challenge Fund, National Natural Science Foundation of China grant 61571301, New York University Global Seed Grants for Collaborative Research, and the Thousand Talents Program for Distinguished Young Scholars  
\end{acknowledgments}

%%%%%%%%%%%%%%%%%%%%%%%%%%%%%%%%%%%%%%%%%%%%%%%%%%%%%%%%%%%%%%%%%%%%%%%%%%%%%%%%%%%%%%%%%%%%%%%%%%%
% How to do the references:
%% 1) First uncomment the below and compile
%\bibliographystyle{apsrev}
%\bibliography{C:/Users/Phymbas/Documents/Bibiliography/NIIReferences}
%% 2) Copy the .bbl file to below and comment out the above two lines.

\end{document}